\documentclass{aastex}          
\usepackage{spr-astr-addons}    

\begin{document}
%
\title{Spatial Periodicity of Galaxy Number Counts, CMB Anisotropy, and SNIa Hubble Diagram Based on the Universe Accompanied by a Non-Minimally Coupled Scalar Field}

\shorttitle{Spatial Periodicity of Galaxy Number Counts}
\shortauthors{Hirano et al.}

\author{Koichi Hirano}
\email{h-koichi@guitar.ocn.ne.jp} 
\and 
\author{Kiyoshi Kawabata}
\affil{Department of Physics, Tokyo University of Science, \\ Shinjuku-ku, Tokyo 162-8601, Japan}
\and 
\author{Zen Komiya}
\affil{Shinjuku College of Information Technology, \\
Nakano-ku, Tokyo 164-0001, Japan}
\affil{Department of Physics, Tokyo University of Science, \\
Shinjuku-ku, Tokyo 162-8601, Japan}


\begin{abstract}
We have succeeded in establishing a cosmological model with a non-minimally coupled scalar field $\phi$ that can account not only for the spatial periodicity or the {\it picket-fence structure} exhibited by the galaxy $N$-$z$ relation of the 2dF survey but also for the spatial power spectrum of the cosmic microwave background radiation (CMB) temperature anisotropy observed by the WMAP satellite.
The Hubble diagram of our model also compares well with the observation of Type Ia supernovae.
The scalar field of our model universe starts from an extremely small value at around the nucleosynthesis epoch, 
remains in that state for sufficiently long periods, allowing sufficient time for the CMB temperature anisotropy to form, and then starts to grow in magnitude at the redshift $z$ of $\sim 1$, followed by a damping oscillation which is required 
to reproduce the observed picket-fence structure of the $N$-$z$ relation. 
To realize such behavior of the scalar field, we have found it necessary to introduce a new form of potential $V(\phi)\propto \phi^2\exp(-q\phi^2)$, with $q$ being a constant. Through this parameter $q$, we can control the epoch at which the scalar field starts growing.
\end{abstract}


\section{Introduction}
It is currently believed, based on observations \citep{rie98,perl99,rie04,ast06,rie07} of a number of Type Ia supernovae (hereafter referred to as SNIa), that the expansion of our universe at present is in an accelerated stage.
Furthermore, the measurements of the temperature anisotropy of the cosmic microwave background radiation (hereafter CMB) by the Wilkinson microwave anisotropy probe (WMAP) \citep{spe07,hin07} have indicated that our universe is flat. 
The high precision of these and other observations have enabled us to intercompare and verify various candidate cosmological models.\par
A flat ${\mit\Lambda}$CDM model consisting of Einstein's cosmological constant $\mit\Lambda$ and cold dark matter, in addition to the ordinary baryonic matter and radiation is currently the standard model because it can account reasonably well for the acceleration of the cosmic expansion as indicated by the SNIa's, the spatial power spectrum of the CMB temperature anisotropy observed by WMAP satellite, and the primordial abundance of the light chemical elements produced through big-bang nucleosynthesis.\par
Nevertheless, some key questions remain as to the validity of the standard ${\mit\Lambda}$CDM model. For instance, the physical origin of the cosmological constant is presently unknown, although it might possibly be a component of {\it vacuum energy}. As such, we have yet to understand why the energy density of the cosmological constant remains unchanged with time. 
 The numerical size of the $\mit\Lambda$-term estimated based on a particle physics consideration is generally more than 120 orders of magnitude too large \citep[see, e.g.,][p.16]{rich01} to be compared to the observed values.
 As an alternative approach to circumvent these problems of the $\mit\Lambda$-term, cosmological models possessing quintessence or a phantom have been proposed \citep[see, e.g.,][]{rat88,pee88,cal02,cal03}. These methods interpret the dark energy in terms of the type of time-varying scalar field.\par
The second consideration is that the cosmic age of the standard model, which is largely based on the WMAP 3-yr data, is supposed to be $13.7$ Gyrs, whereas the possibility of it being, for instance, approximately $14\sim 15$ Gyrs has been raised based on observations of certain globular clusters.
We have already shown that we might be able to avoid this cosmic age problem by introducing a time-decaying constant as well as the effect of galaxy merger with  a time scale of approximately 3 Gyrs \citep{kom05,kom06}. However, in view of the various uncertainties involved in cosmic age determination processes,  $13$ Gyrs appears to be a conservative estimate for the lowerbound for any cosmological modeling, as indicated by \citet{kom06}.\par
The third and most baffling problem, however, is the spatial periodicity of approximately $128 h_{100}^{-1}$ Mpc with $h_{100}\equiv H_0~{\rm km~s^{-1} Mpc^{-1}} /(100$ ${\rm km~s^{-1} Mpc^{-1}}$) emerging in the relation of the galaxy number counts $N$ and the redshifts $z$ relation (hereafter referred to as the {\it picket-fence structure} of the $N$-$z$ relation) originally discovered by \citet{bro90} through a pencil-beam survey of field galaxies. In each of the four directions of the survey, an epoch of relatively high galaxy number counts and another  that exhibits comparatively low number counts appear alternately, as though concentric spherical shells (or epochs) of higher galaxy number densities were surrounding us with their individual centers situated at our location.\par
Such an observed characteristic feature, if it were reflecting an actual physical concentration of galaxies existing at certain distances from us, would definitely be incompatible with the cosmological principle that presumes the uniformity and isotropy of our space-time.
In fact, based on a large number of numerical simulations using the Einstein-de Sitter model as well as the ${\mit\Lambda}$CDM models, \citet{yos01}, among others, demonstrated that the probability to obtain a periodic structure in the spatial distribution of galaxies such as that inferred from the results of \citet{bro90} should be less than 1/1000. Hence, it is often the case that the existence of the picket-fence structure reported by \citet{bro90} is dismissed as an observational fluke. \par 
However, this problem is not so simple. The $N$-$z$ relation obtained from the 2dF Galaxy Redshift Survey (2dF GRS) also exhibits a definite picket-fence structure at the interval of roughly $\Delta z=0.03$ \citep[see, e.g., Fig. 17 of][]{col01}.
 The locations of higher galaxy number counts are found at $z \simeq 0.03$, $0.06$, $0.08$, $0.11$, $\dots$, which is essentially the same as those found by \citet{bro90}.
Even in the $N$-$z$ relation obtained through the Sloan Digital Sky Survey (hereafter  SDSS), protrusions still appear in the locations of relatively higher galaxy number counts compared with those predicted from the standard model (see Fig. 2 of \citet{teg04}) at $z \simeq 0.03$, $0.05$, $0.08$, $0.11$, $\dots$, although their amplitudes of variation are much less pronounced compared to those observed in the 2dF GRS data. \par
For the quantitative comparison of our theoretical computations, we would like to adopt the $N$-$z$ relation taken from the 2dF GRS data for the following reasons:
i) the sky coverage of the 2dF data is more symmetric with respect to the celestial equator than that of the SDSS data, including not only a major portion of the Southern hemisphere sky, but also a comparable portion  of the Northern hemisphere sky, whereas the renderings of the SDSS project described by \citet{ade07} indicates a notable asymmetry in the hemispheric coverages, and ii)the width of the bin for number count statistics used for the 2dF data is narrower by a factor of approximately 2 compared with that  for the SDSS data. \par
One viable explanation for the picket-fence structure of the $N$-$z$ relation 
 within the framework of the Robertson-Walker metric would be the introduction of a scalar field that non-minimally couples with the curvature scalar, as proposed by \citet{mor90,mor91}. Morikawa's model was subsequently extended by \citet{kas94}, and later by \citet{fuk97}, to include the effects of matter and the $\mit\Lambda$ term.  Such coupling can produce the picket-fence structure in the following  way: the scalar field model of \citet{mor90,mor91} assumes a potential $V(\phi)$ proportional to $\phi^2$ so that the state of our universe oscillates back and forth around the potential minimum, inducing an oscillation of the Hubble parameter as well, due to the curvature coupling of $\phi$. Hence, an epoch of relatively rapid expansion and an epoch of relatively slower expansion appear alternately due to this oscillation. 
 During the former epoch, the number density of galaxies would diminish noticeably; whereas during the latter epoch, the decrease in the number density would remain less prominent. We would then get a picket-fence structure as an apparent or illusionary effect in the $N$-$z$ relation, because we are observing the two types of epochs alternately with increasing value of $z$. \par
  Since the spatial distribution of galaxies remains homogeneous for a given instance of time, the cosmological principle can still be preserved. 
  Note that \citet{fuk97} pointed out that the adoption of negative values for the coupling constant $\xi$ yields a succession of mini-inflationary (concave shaped) cosmic expansion with time, thereby prolonging the cosmological age. We shall therefore employ negative values for $\xi$. 
The temporal oscillations of the fundamental constants, such as the gravitational constant $G$, that would be caused by the scalar field \citep{hil90} and the possible presence of a coherent peculiar velocity field are likely to enhance that apparent effect \citep{hil91}.\par
On the other hand, based on the primordial abundances of the light chemical elements, \citet{sal96}, and later \citet{que97}, argued that the scalar field needs to be at an extremely low level during the time periods, including that for the big bang nucleosynthesis, and demonstrated that under certain conditions, it is possible to suppress the scalar field $\phi$ to an extremely small value ($\phi < 10^{-10}$) over certain periods of time at the approximate time of nucleosynthesis, which subsequently starts to grow in magnitude and simultaneously exhibits damping oscillations as we move toward the present.\par
 We shall therefore restrict ourselves to the models in which the scalar fields undergo stationary states with sufficiently small values over the relevant periods of time. One notable problem associated with computing such cosmological models starting from the present epoch, as is usually done, is that one must perform a fine-tuning for the initial value for $\phi$ to a large number of decimal places in order to arrive at the stationary state of $\phi$ at or near the time of big bang nucleosynthesis. However, we have found that by solving the evolution equations from the time of this nucleosynthesis, when $\phi\simeq 0$, toward the present, we need not carry out any such fine-tuning, as was suggested by \citet{sal96}. The theoretical basis and a number of sample computations will be  published elsewhere.\par
In general, we cannot hope to sufficiently constrain the parameter values unless more than one type of observational data are employed, as we ourselves have learned in other investigations \citep{kom06}. Therefore, we also make use of another crucial observation, viz., the spatial power spectrum of the CMB temperature anisotropy obtained from the WMAP 3-yr data \citep{spe07,hin07} to narrow the possible ranges of the model parameters. The candidate model thus obtained will be further cross-checked against the observed Hubble diagram of SNIa's produced with the Gold data set of \citet{rie07}. \par
The purpose of the present study is thus twofold. First, we investigate whether it is possible to determine a cosmological model possessing the non-minimally coupled Morikawa scalar field that can account for both the picket-fence structure of the $N$-$z$ relation of the field galaxies and the power spectrum of the CMB temperature anisotropy given by the WMAP 3-yr data. Second, we attempt to determine a simple method by which to modify the basic equation or equations to construct a model that can better serve our purpose, if the above-mentioned model is found to be inappropriate. We herein demonstrate that the use of a potential for the scalar field that is somewhat different from that employed by \citet{mor90} is better for this purpose.\par
The remainder of the present paper is organized as follows. 
In Section 2, we present the set of evolution equations for our universe, incorporating the newly introduced potential for $\phi$, and discuss the characteristic features of our cosmological model compared to those of the standard $\mit\Lambda$CDM model. In Section 3, we investigate the effect of each of the principal model parameters, including the power index $q$ employed for our newly proposed potential on the theoretical computations of the $N$-$z$ relation and the power spectrum of the CMB temperature anisotropy, and show that there exists a set of parameter values with which we can reproduce, rather satisfactorily, both the 2dF data and the WMAP 3-yr data. In addition, we demonstrate that the model thus obtained can account for the observed Hubble diagram of the recently observed SNIa's, at least to a degree comparable to that of the standard model. In Section 4, we summarize our results and present our conclusions.

\section{Basic Equations for Theoretical Computations}
\subsection{Derivations of Time Evolution Equations}
The relevant Lagrangian $L$ for our cosmological model is as shown below:
\begin{eqnarray}
L & = & \frac{1}{2}\xi R\psi^2-\frac{c^4}{16\pi G}R+\frac{1}{2}g^{\mu\nu}\partial_{\mu}\psi\partial_{\nu}\psi \nonumber \\
  &   & -\frac{1}{2}\left(\frac{m_\phi c}{\hbar}\right)^2\psi^2\exp{\left(-q\frac{4\pi G}{3c^4}\psi^2\right)} \nonumber \\
  &   & +\frac{c^4}{8\pi G}\mit\Lambda +L^{(\rm mr)}, \label{Lagrangean}
\end{eqnarray}
where $g^{\mu\nu}$ are the $(\mu,\nu)$-components of the metric tensor,  $\psi$ is the scalar field, $R$ is the scalar curvature, $\xi$ is the coupling constant, $q$ is a scalar constant, $m_\phi$ is the mass of the scalar field, $G$ is Newton's gravitational constant, $c$ is the velocity of light, $\mit\Lambda$ is Einstein's cosmological constant, and $L^{(\rm mr)}$ is the Lagrangian due to matter and radiation. In addition, we specify that $\partial_{\mu}\psi\equiv \partial\psi/\partial x^{\mu}$. Each Greek subscript ranges from 0 to 3, and Einstein's summation convention is implicitly assumed in the above equation, as well as the equations presented hereafter. 
Then, based on the principle of least action, we extremize the action integral  
\begin{equation}
S=\int d^4 x \sqrt{-g}\ L
\end{equation}
by setting $\delta S=0$, which leads us to the following gravitational field equations: 
\begin{eqnarray}
\lefteqn{R_{\mu\nu}-\frac{1}{2}g_{\mu\nu}R+\mit\Lambda g_{\mu\nu}} \nonumber \\
& = &\frac{8\pi G}{c^4}\Bigg[\partial_{\mu}\psi\partial_{\nu}\psi-\frac{1}{2}g_{\mu\nu}\partial_{\rho}\psi\partial^{\rho}\psi  \Bigg. \nonumber \\
&   &+\xi g_{\mu\nu}\square \psi^2-\xi[\psi^2]_{;\mu\nu}+\xi\psi^2\left(R_{\mu\nu}-\frac{1}{2}g_{\mu\nu}R\right) \nonumber \\
&   &+\frac{1}{2}g_{\mu\nu}\left(\frac{m_\phi c}{\hbar}\right)^2 \psi^2\exp{\left(-q\frac{4\pi G}{3c^4}\psi^2\right)} \nonumber \\
&   & \Bigg.+T^{(\rm mr)}_{\mu\nu} \Bigg], \label{graveq}
\end{eqnarray}
where the symbol~"$\square$" signifies the d'Alembertian, the symbol~"$(\cdots)_{;\mu}$" is the covariant derivative of a quantity "$\cdots$" with respect to spatial coordinate $x^\mu$, and the quantities 
$T^{(\rm mr)}_{\mu\nu}$ are the $(\mu,\nu)$-components of the energy-momentum tensor of  matter and radiation.\par
Note that the potential term (the fourth term on the right-hand side of Eq.(\ref{Lagrangean})) reflects our modification to the original form employed by \citet{mor90,mor91}, 
which coincides with the latter if $q=0$. As we shall show later herein, this new form of potential enables us to control the epoch when the growth of the scalar field $\psi$ begins to take place. The closer this epoch is to the present era, the less 
affected are the amplitudes of the spatial power spectrum of the CMB temperature anisotropy in the large-scale domain because of the reduced  effect  of  the late-time integrated Sachs-Wolfe effect \citep[see, e.g.,][]{nas06}. Based on \citet{kas94}, we have included the $\mit\Lambda$ term, viz., the fifth term on the right-hand side of Eq.(\ref{Lagrangean}) because it provides greater flexibility,  yields a better fit to the CMB spectrum, and makes the age of our universe older by dominating the time periods prior to the resurgence of the scalar field $\phi$. The fact that we have a stationary state of $\phi$ at an extremely small value does not necessarily mean that the universe was born with $\phi\simeq 0$. On the contrary, the Big Bang is supposed to have been accompanied by a scalar field of considerable magnitude, as has been pointed out by \citet{fuk97}.\par 
In accordance with the inflationary cosmology, we restrict ourselves to 
a flat geometry, so that the Robertson-Walker line element ${\rm d}s$ takes the following form:
\begin{equation}
{\rm d}s^2=c^2{\rm d}t^2-a^2(t)[{\rm d}\chi^2+\chi^2({\rm d}\theta^2+\sin^2\theta~{\rm d}\phi^2)].
\end{equation}
Here, $a(t)$ is the cosmic scale factor at proper time $t$ normalized to the present-day value, whereas $\chi$, $\theta$, and $\phi$ are the comoving angular coordinates.
The matter included in our model universe is assumed to consist of baryonic matter as well as cold dark matter (CDM), but not massive neutrinos.
Furthermore, the scalar field is assumed to interact with the gravitation
 merely by coupling with the scalar curvature, so that there is no mutual energy exchange between the matter, radiation, and scalar field. Under this assumption, the equation of state for each constituent is given as follows:
\begin{eqnarray}
\dot{\rho_m}c^2&=&3\frac{\dot{a}}{a}(\rho_m c^2+P_m)=0,
\\
\dot{\rho_r}c^2&=&3\frac{\dot{a}}{a}(\rho_r c^2+P_r)=0, \\
\dot{\rho_{\psi}}c^2&=&3\frac{\dot{a}}{a}(\rho_{\psi} c^2+P_{\psi})=0, \label{eqstatepsi}
\end{eqnarray}
together with 
\begin{equation}
 P_m = 0 ;  \quad  P_r = \frac{1}{3}\rho_r c^2,
\end{equation}
where the quantities $\rho$ and $P$ stand for the mass density and the pressure of each constituent of the universe, respectively, and the subscripts m, r, and $\psi$ designate ''matter" (baryon+CDM), ''radiation", and ''scalar field", respectively. We also employ a short-hand notation $\dot{A}\equiv {\rm d}A/{\rm d}t_{\rm c}$,
where $t_{\rm c}$ is the conformal time defined as
\begin{equation}
 t_{\rm c} = \int\frac{c}{a}{\rm d}t,~~~~{\rm so~that}~~ 
\dot{A}=\frac{{\rm d}A}{{\rm d}t_c}=\frac{a}{c}\frac{{\rm d}A}{{\rm d}t}. 
\end{equation}
\par
On the other hand, the mass density $\rho_\psi$ and the pressure $P_\psi$ for the scalar field $\psi$ resulting from the principle of least action are as shown below:
\begin{eqnarray}
\rho_{\psi}c^2 & = & \frac{1}{2}\frac{1}{a^2}(\dot{\psi})^2+3\xi\frac{(\dot{a})^2}{a^4}\psi^2+6\xi\frac{\dot{a}}{a^3}\psi\dot{\psi} \nonumber \\
               &   & +\frac{1}{2}\left(\frac{m_\phi c}{\hbar}\right)^2\psi^2\exp{\left(-q\frac{4\pi G}{3c^4}\psi^2\right)}, \label{rhopsi}
\end{eqnarray}
\begin{eqnarray}
P_{\psi}&=&\frac{1}{2}\frac{1}{a^2}(\dot{\psi})^2-2\xi\frac{\ddot{a}}{a^3}\psi^2+\xi\frac{(\dot{a})^2}{a^4}\psi^2  \nonumber \\
    & & -2\xi\frac{1}{a^2}\psi\ddot{\psi}-2\xi\frac{1}{a^2}(\dot{\psi})^2-2\xi\frac{\dot{a}}{a^3}\psi\dot{\psi} \nonumber \\
    & & -\frac{1}{2}\left(\frac{m_\phi c}{\hbar}\right)^2\psi^2\exp{\left(-q\frac{4\pi G}{3c^4}\psi^2\right)}.  \label{Ppsi}
\end{eqnarray}
Substituting Eqs.(\ref{rhopsi}) and (\ref{Ppsi}) into Eq.(\ref{eqstatepsi}), we obtain the time evolution equation for the scalar field:
\begin{equation}
\ddot{\phi}=-2\frac{\dot{a}}{a}\dot{\phi}-6\xi\frac{\ddot{a}}{a}\phi-a^2\left(\frac{m_\phi c}{\hbar}\right)^2(1-q\phi^2)\phi\exp{(-q\phi^2)}, \label{EqPhidots}
\end{equation}
where we have employed a dimensionless quantity $\phi$, instead of $\psi$:
\begin{equation}
\displaystyle{ \phi=\sqrt{{4\pi G\over 3c^4}}~\psi}. \label{eq:psiphi}
 \end{equation}
From the (0,0)-components and (1,1)-components of Eq.(\ref{graveq}), and using Eq.(\ref{EqPhidots}), we obtain the following set of the time evolution equations for the cosmic scale factor $a$:
\begin{eqnarray}
\frac{\dot{a}}{a} & = & \Bigg[6\xi\phi\dot{\phi}+\left\{(6\xi\phi\dot{\phi})^2+(1-6\xi\phi^2) \  \right. \nonumber \\
                  &   & \times\left((\dot{\phi})^2+a^2\left(\frac{m_\phi c}{\hbar}\right)^2\phi^2\exp(-q\phi^2) \right.\nonumber \\
                  &   & \left.\left.+\frac{H_0^2}{c^2}\left(\frac{\Omega_{m,0}}{a}+\frac{\Omega_{r,0}}{a^2}+\Omega_{{\mit\Lambda},0}a^2\right)\right)\right\}^{1/2}\Bigg] \nonumber \\
                  &   & / (1-6\xi\phi^2),     \label{EqAdot}
\end{eqnarray}
\begin{eqnarray}
\frac{\ddot{a}}{a} & = & \Bigg[2\frac{\dot{a}^2}{a^2}(1-6\xi\phi^2)-3(\dot{\phi})^2-24\xi\frac{\dot{a}}{a}\phi\dot{\phi}+6\xi(\dot{\phi})^2 \nonumber \\
                   &   & -6\xi a^2\left(\frac{m_\phi c}{\hbar}\right)^2(1-q\phi^2)\phi^2\exp{(-q\phi^2)} \nonumber \\
                   &   & -\frac{3}{2}\frac{H_0^2}{c^2}\left(\frac{\Omega_{m,0}}{a}+\frac{\Omega_{r,0}}{a^2}\right)\Bigg] \nonumber \\
                   &   & / \{1-6\xi\phi^2(1-6\xi)\},   \label{EqAdots}
\end{eqnarray}
where $H_0$ is the Hubble constant. The quantities $\Omega_{\rm m,0}$, $\Omega_{\rm r,0}$, and $\Omega_{{\mit\Lambda},0}$ are the present mass densities of matter, radiation, and the cosmological constant $\mit\Lambda$ normalized, respectively, to the present value of the critical density $\rho_{\rm c,0}(=3H_0^2/8\pi G)$, viz., $\Omega_{\rm m,0}\equiv\rho_{\rm m,0}/\rho_{\rm c,0}$, $\Omega_{\rm r,0}\equiv\rho_{\rm r,0}/\rho_{\rm c,0}$ and $\Omega_{{\mit\Lambda},0}=(c/H_0)^2{\mit\Lambda}/3$. The density parameter of the scalar field $\psi$ for the present epoch is similarly defined as $\Omega_{\psi,0}\equiv\rho_{\psi,0}/\rho_{\rm c,0}$. Since the mass density $\rho_\phi$ in terms of $\phi$ should be equal to $\rho_\psi$, we have $\Omega_{\phi,0}=\Omega_{\psi,0}$. Furthermore, since we are concerned only with the flat geometry case, we obtain the following constraint from Eq.(\ref{EqAdot}) for the present era:
\begin{equation}
\Omega_{m,0}+\Omega_{r,0}+\Omega_{\phi,0}+\Omega_{{\mit\Lambda},0} = 1,
\end{equation}
so that 
\begin{equation}
\Omega_{{\mit\Lambda},0}=1-\left(\Omega_{m,0}+\Omega_{r,0}+\Omega_{\phi,0}\right),
\end{equation}
which is to be substituted to Eq.(\ref{EqAdot}).
\\
\subsection{Stationary State and Onset of Damping Oscillation of Scalar Field}
\citet{sal96} and \citet{que97} indicated that the magnitude of the scalar field must remain extremely small  during the epochs around the big bang nucleosynthesis  in order not to spoil the successful prediction made on the cosmic abundances of light elements based on the Einstein-de Sitter model universe.
However, to give rise to such a plateau state of the scalar field $\phi$ at those   periods of time without undertaking any extreme fine-tuning of the presentday value of $\phi_0$ such as those carried out in \citet{sal96} and \citet{que97}, we 
 would like to start integrating the evolution equations at the time  $t_{\rm c,init}$ corresponding to $a=10^{-9}$, which is sufficiently close to the big bang nucleosynthesis toward the present time $t_{\rm c,0}$. \par 
In view of the fact that the universe at this stage is dominated by radiation, we approximate the values of $a(t_{\rm c, init})$ and $\dot{a}(t_{\rm c, init})$ with those given by the ${\mit\Lambda}$CDM model for a given set of the values of $\Omega_{\rm m,0}$, $\Omega_{\rm r,0}$, and $\Omega_{\Lambda,0}$.
Furthermore, we take $\dot{\phi}(t_{\rm c, init})=0$ because of the assumed stationary state of $\phi$.
Then, $\phi(t_{\rm c,init})$ is the only parameter that needs to be specified at $t_{\rm c, init}$, which is determined in the following manner: 
a starting value for $\phi(t_{\rm c, init})$ is picked, and the model computation is carried out all the way to the present time, and the resulting value of the Hubble parameter $H_0$ is inspected to see if it agrees with 72 ${\rm km~s^{-1} Mpc^{-1}}$, the intended value,  within 0.1\%. If not, we multiply $\phi(t_{\rm c, init})$ by 10, and perform  the model computation again. 
This procedure is repeated until we  bracket 72 ${\rm km~s^{-1} Mpc^{-1}}$, after which the Newton method approximating $H_0-72$ with a piece-wise linear function of $\log \phi_{\rm init}$ is applied to find the desired value of  $\phi(t_{\rm c, init})$.\par
A sample behavior of the scalar field $\phi$ is shown in Fig. 1 as a function of scale factor $a$ for a model universe with  Morikawa scalar field model($q=0$) obtained using the procedure described above: we have employed $\Omega_{\rm m,0}=0.237$, $\Omega_{\rm r, 0}=8.066\times 10^{-5}$ corresponding to $T_{\rm CMB,0}=2.726$K, $\xi=-40$, and $m_{\rm s}=3.7\times 10^{-31}h_{100}$~eV.    Note that cosmic age $t_{\rm init}=23.86$ s (yielding $t_{\rm c, init}=4.63\times 10^{-4}$ Mpc), $\dot{a}(t_{\rm c, init})=2.157\times 10^{-6}$, and $\phi(t_{\rm c, init})=2.341\times 10^{-30}$ at $a=10^{-9}$, from which $H_0=72.031$ ${\rm km~s^{-1} Mpc^{-1}}$ and $\phi_0=7.979\times 10^{-5}$ result.  

\begin{figure}[h!]
\includegraphics[scale=0.43]{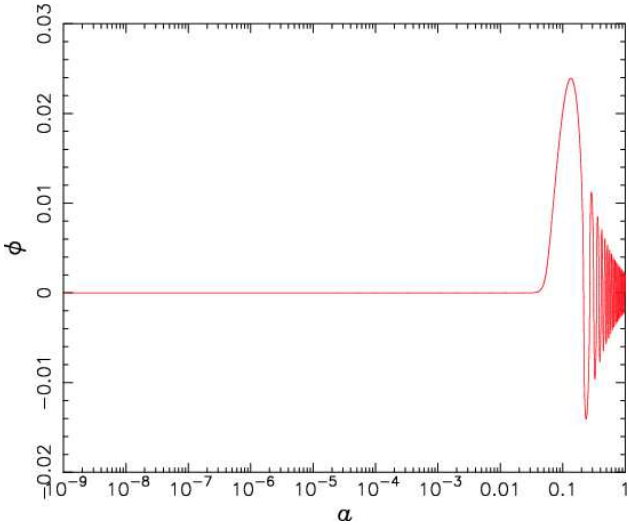}
\caption{An example of the variation of the Morikawa-type  scalar field $\phi$ as a function of scale factor $a$, starting from a stationary state at $a=10^{-9}$(see the text for the model description).   
\label{fig:salgado_m}}
\end{figure}
\begin{figure}[h!]
\includegraphics[scale=1.06]{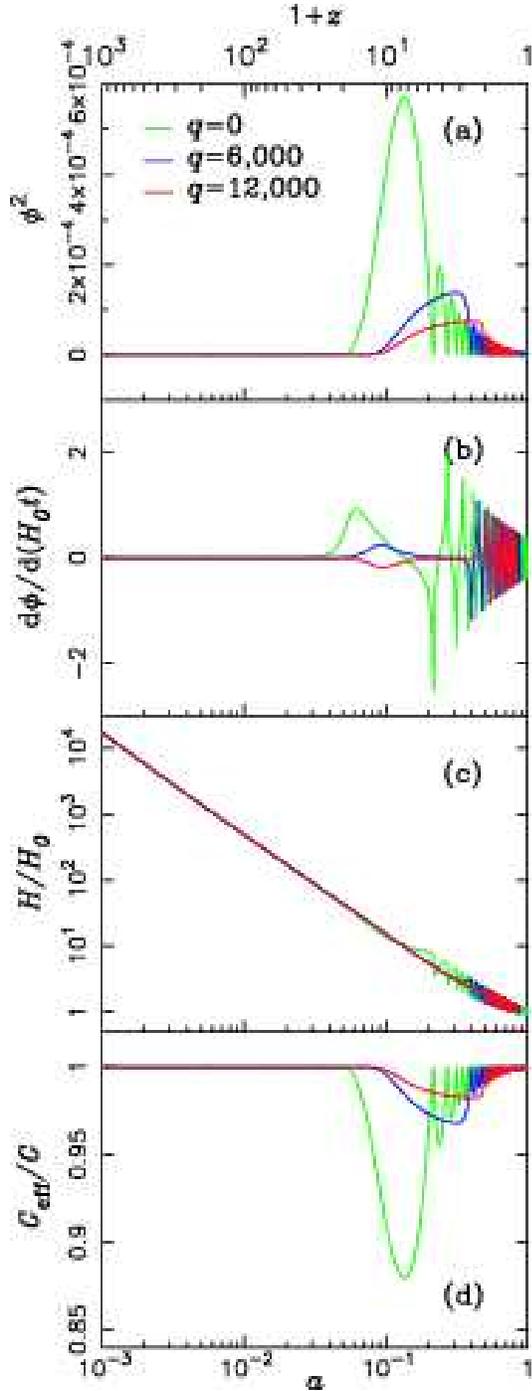}
\caption{ (a) Dependence of $\phi^2$ on the value of $q$. The abscissa is the scale factor $a$; 
(b) Dependence on $q$ of the derivative of $\phi$ with respect to Hubble time $H_0t$. The abscissa is the scale factor $a$;
(c) Dependence on $q$ of the onset time of the oscillation of the Hubble parameter $H/H_0$ normalized to the Hubble constant. The abscissa is the scale factor $a$;
(d) Dependence on $q$ of the time for the onset of oscillation of the effective gravitational constant $G_{\rm eff}/G$ normalized to the Newton's gravitational constant $G$. The abscissa is the scale factor $a$. \label{fig:model}}
\end{figure}
\begin{figure}[h!]
\includegraphics[scale=0.34]{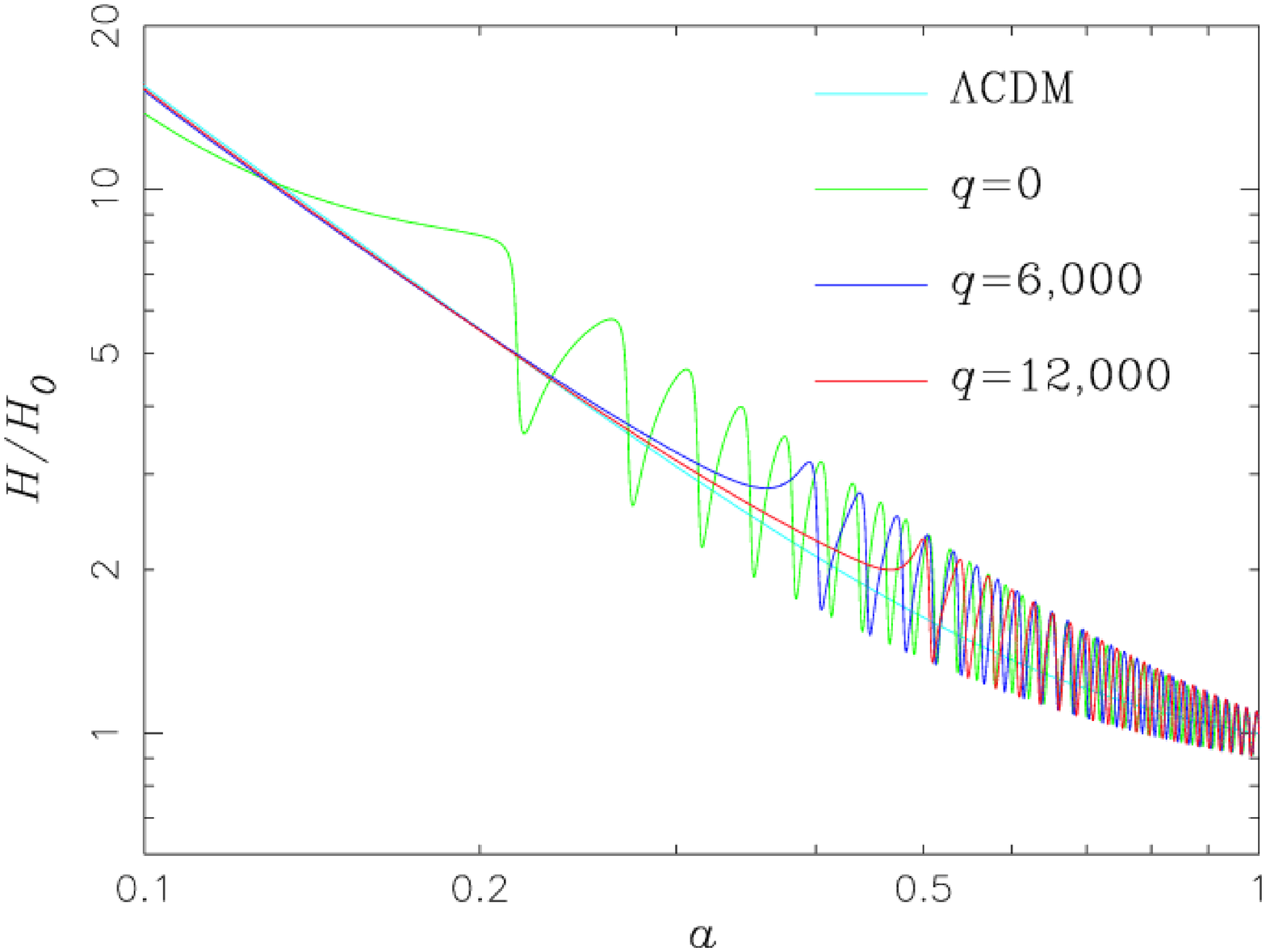}
\caption{Detail of the behavior of $H/H_0$ depicted in Fig.\ref{fig:model}(c) near the recent epochs.
 \label{fig:h_dai}}
\end{figure}
\par
Next, we want to demonstrate that the time for $\phi$ to start growing can be 
controlled by changing the value of the power index $q$ introduced in Eq.(\ref{EqAdot}). 
Figure \ref{fig:model}(a) shows the variation of $\phi^2$ as a function of scale factors $a$ for three values of $q$, viz., $0$, $6,000$, and $12,000$.
The values of the model parameters employed are as follows: $\Omega_{\rm m,0}=0.237$, $\Omega_{\phi,0}=0.15$(hence $\Omega_{{\mit\Lambda},0}=0.613$),
 $h_{100}=0.72$, $\xi=-40$, and $m_\phi=3.7\times 10^{-31}h_{100}~{\rm eV}$. They  are taken from  our best-candidate model deduced based on a large number of trial and error search in comparison with the observational data of the galaxy $N$-$z$ relation and the spatial power spectrum of the CMB, which we shall discuss shortly. Note that the onset of the damping oscillation of $\phi^2$ is delayed as the value of $q$ increases. 
Figs.\ref{fig:model}(b), \ref{fig:model}(c), and \ref{fig:model}(d) indicate 
 the corresponding  derivative of $\phi$ with respect to the Hubble time $H_0t$, Hubble parameter $H/H_0$(see also Fig.\ref{fig:h_dai}, an close-up view of this diagram), and effective gravitational constant $G_{\rm eff}$ given by 
\begin{equation}
\frac{G_{\rm eff}}{G}=\frac{1}{1-6\xi\phi^2},  \label{EqGeff}
\end{equation}
respectively, where $G$ is Newton's gravitational constant \citep{mor91}\footnote{The scalar field  $\psi$ used in the present work corresponds to $\phi$ in \citet{mor90,mor91}. Hence, the quantity $\phi$ used in the present study is equal to $\sqrt{4\pi G/3c^4}$ times Morikawa's $\phi$. See also Eq.(\ref{eq:psiphi}) of the present work.}. These quantities exhibit similar oscillatory behaviors
 as functions of $a$ to that of $\phi^2$.
\section{Observational Tests}
\subsection{Galaxy Number Counts with Respect to Redshift}
As has been mentioned earlier, the observed data on the galaxy number counts as a function of redshift $z$ obtained through the 2dF survey \citep[see. e.g., Fig. 17 of][]{col01} exhibit alternating regions of comparatively high counts and comparatively low counts at the interval of approximately $\Delta z=0.03$, essentially the same feature as that discovered by \citet{bro90} through their pencil-beam survey, which is referred to herein as the {\it picket-fence structure}. 
It should be emphasized that the standard model, namely the flat $\mit\Lambda$CDM model derived on the basis of the WMAP observations, is not compatible, or is at least extremely difficult to reconcile with this picket-fence structure, unless we are willing to set aside the cosmological principle. \par   
 In this section, we compare the $N$-$z$ relationships for the ${\rm b_J}$-band computed using the proposed model having the new potential for the scalar field  and the 2dF survey data \citep{col01}. 
 The theoretical computations are carried out using the computer code {\it ncmod} \citep{gar98} with a modification to incorporate our cosmological model.  Briefly speaking, we adopt the following values for the \citet{sch76} parameterization derived by \citet{nor02} based on the fitting 
 to the 2dF data: $M^{\star}_{\rm b_J}-5\log_{10}h_{100}=-19.66, ~\alpha=-1.21, ~\Phi^{\star}=1.61\times 10^{-2}h_{100}^3 {\rm Mpc}^{-3}$. 
 Here, it is assumed that galaxies fainter than $m_{\rm b_J}$=$19.45$ cannot be seen, which is taken into account in our computations in terms of the selection function. 
 Furthermore, the spectral energy distributions (SED) of different types of galaxies are those employed by \citet{yos88}. 
\par
\begin{figure}[h!]
\includegraphics[scale=0.37]{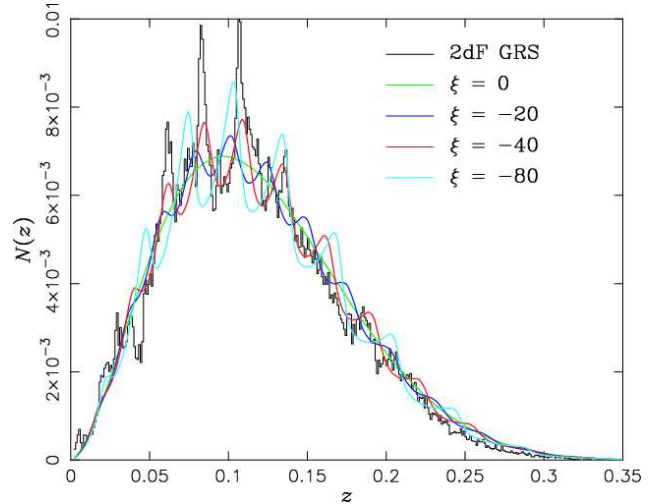}
\caption{Dependence on the coupling constant $\xi$ of the computed $N$-$z$ relation of galaxies. The abscissa indicates the redshift $z$, while the ordinate is the fractional number count $N$. For comparison, the 2dF data are also plotted. \label{fig:nz_xi}}
\end{figure}
First, let us investigate the dependence of the theoretical $N$-$z$ relationship on each of the model parameters.
Figure \ref{fig:nz_xi} shows the theoretical $N$-$z$ relationships computed 
 for $\xi= 0$, $-20$, $-40$, and $-80$ using the proposed model with the new potential. The values of the other model parameters are as follows: $\Omega_{\rm m,0}=0.237$, $\Omega_{\phi,0}=0.15$ (hence $\Omega_{{\mit\Lambda},0}=0.613$), $h_{100}=0.72$, $m_\phi=3.7\times 10^{-31}h_{100}~{\rm eV}$, and $q=12,000$, which are taken from our best-fit model (see Fig.\ref{fig:cmb_q_m}), which is capable of reproducing both the 2dF data and the WMAP 3-yr data. 
Note that no spatial periodicity arises for $\xi=0$, but that the amplitude of variation of the galaxy number count $N$ with $z$ tends to be enhanced as the absolute value of $\xi$ is increased. In addition, a slight change in the period of oscillation takes place with the value of $\xi$ as well. As a result, $\xi=-40$ is selected. \par
\begin{figure}[h!]
\includegraphics[scale=0.37]{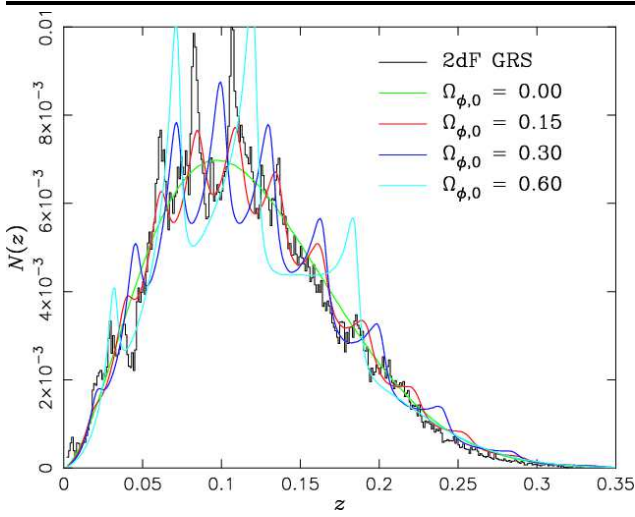}
\caption{Dependence on $\Omega_{\phi,0}$ of the computed $N$-$z$ relationship of galaxies. The abscissa indicates the redshift $z$, while the ordinate is the fractional number count $N$. For comparison, the 2dF data are also plotted.
 \label{fig:nz_omegaphi}}
\end{figure}
\begin{figure}[h!]
\includegraphics[scale=0.37]{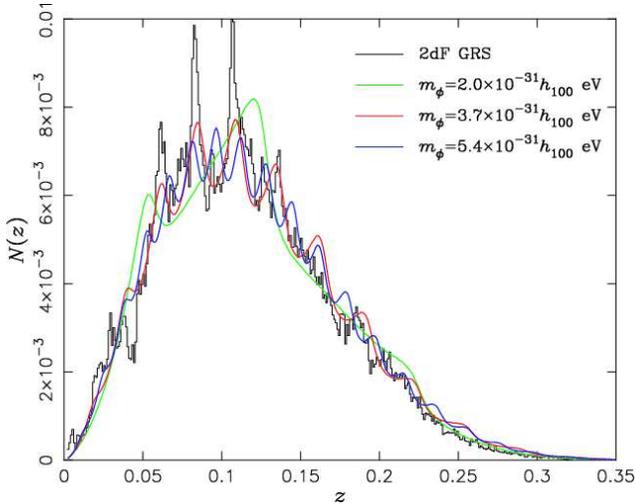}
\caption{Dependence on the mass $m_\phi$ of the scalar field, of the computed $N$-$z$ relationship of galaxies. The abscissa indicates the redshift $z$, while the ordinate is the fractional number count $N$. For comparison, the 2dF data are also plotted. \label{fig:nz_ms}}
\end{figure}
Figure \ref{fig:nz_omegaphi} shows the dependence of the $N$-$z$ relationship on the density parameter of the scalar field  $\Omega_{\phi,0}$, for which we have employed 0.00, 0.15, 0.30, and 0.60 together with $\xi=-40$, whereas the values of the remaining parameters are kept the same as those employed for Fig. \ref{fig:nz_xi}.  Note that the amplitude of variation of $N$ increases with increasing $\Omega_{\phi,0}$. The case in which $\Omega_{\phi,0}=0$ is found to be essentially the same as that for the flat $\mit\Lambda$CDM model with $\Omega_{m,0}=0.237$ because the value of $\phi$ remains sufficiently small throughout the relevant epochs.  We select $\Omega_{\phi,0}=0.15$ as the best choice. \par
Figure \ref{fig:nz_ms} shows the theoretical $N$-$z$ relationships for $m_\phi=2.0\times 10^{-31}h_{100}~{\rm eV}$, $3.7\times 10^{-31}h_{100}~{\rm eV}$, and $5.4\times 10^{-31}h_{100}~{\rm eV}$, respectively.
The values of the remaining parameters are the same as those for Fig. \ref{fig:nz_omegaphi}, except that we have applied $\Omega_{\phi,0}=0.15$.
The period of oscillation of $N$ with respect to the redshift $z$ tends to decrease and its amplitude becomes smaller with increasing $m_\phi$. We select $3.7\times 10^{-31}h_{100}$ as the best value for $m_\phi$. \par  
\begin{figure}[h!]
\includegraphics[scale=0.37]{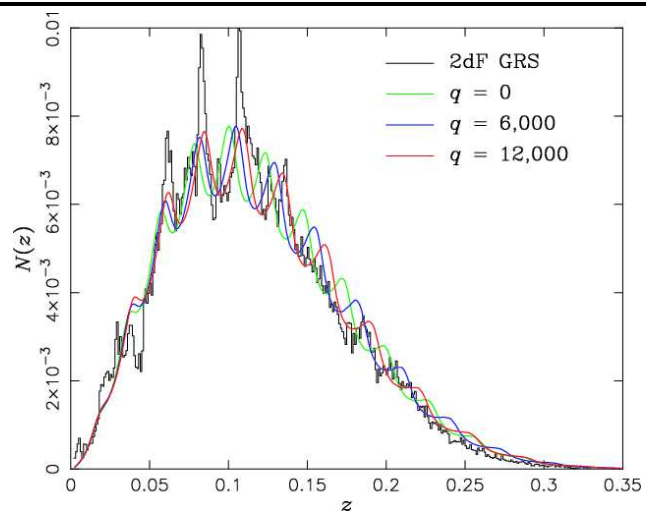}
\caption{Dependence on the power index $q$ of the computed $N$-$z$ relationship of galaxies. The abscissa indicates the redshift $z$, while the ordinate is the fractional number count $N$.   \label{fig:nz_q}}
\end{figure}
Figure \ref{fig:nz_q} depicts the theoretical $N$-$z$ relationships computed for  $q=0$, $6,000$, and $12,000$, respectively. The values of the other parameters are the same as those used for Fig.\ref{fig:nz_ms}, except that $m_\phi=3.7\times 10^{-31}h_{100}$ is used.
 Note that the amplitude of variation of $N$ is only slightly affected by the value of $q$, although the period of oscillation changes slightly. In other words, we could maintain the amplitudes of $N$ variation at sufficient magnitudes even if we vary the value of $q$ over an appreciable range.  
Nevertheless, in the next section, we shall show that the value of $q$ significantly affects the large-scale portion of the spatial power spectrum of the CMB temperature anisotropy, which enables us to select 12,000 as the best value for this parameter, and arrive at a model that can account for both the picket-fence structure of the $N$-$z$ relationship of galaxies and the observed power spectrum of the CMB temperature anisotropy. 
\par
Note that the $\mit\Lambda$CDM model predicts only a smooth $N$-$z$ curve without abrupt changes, just like the case of $\Omega_{\phi,0}=0$ in Fig.\ref{fig:nz_omegaphi}.
Unlike the cases for the original Morikawa models \citep{mor90,mor91,fuk97}, the scalar field in the present models begins to exhibit significant oscillation of appreciable amplitude at approximately $z\sim 1$, but is rapidly damped toward the present. Note that as the value of $q$ increases, the onset of the oscillation shifts toward a more recent epoch. However, since we observe the galaxies only up to $z=0.35$ or so in the 2dF data, an exact choice of the $q$ value is not important with respect to the theoretical $N$-$z$ relationship. However, later herein, we shall demonstrate the fundamental importance of the non-zero $q$ in simultaneously producing any satisfactory fit of the theoretical spectrum to the WMAP data on the CMB temperature anisotropy.  \par
\subsection{Cosmic Microwave Background Radiation}
The high-resolution measurement of the spatial power spectrum of the temperature anisotropy of the CMB by the Wilkinson microwave anisotropy probe (WMAP) should provide us with another crucial test for the proposed model. 
For this purpose, we employ the WMAP 3-yr project data \citep{spe07,hin07}. \par
For the set of our cosmological models that have provided reasonable fits to the observed galaxy $N$-$z$ relation, we calculate the theoretical angular power spectrum of the CMB temperature anisotropy, while making use of a modified Boltzmann code CAMB \citep{lew00}, which is based on a line-of-sight integration approach, as employed in the CMBFAST code \citep{sel96}. We first tabulate the values of \{$a$, $\dot{a}$, $\ddot{a}$, $\phi$, $\dot{\phi}$, and $\ddot{\phi}$\} obtained through integrating Eqs.(\ref{EqPhidots}), (\ref{EqAdot}),  and (\ref{EqAdots}) with respect to the conformal time.
 These tables are then used to perform interpolations to find the values of these variables for a given time inside the power spectrum computation program. \par
\begin{figure}[h!]
\includegraphics[scale=0.43]{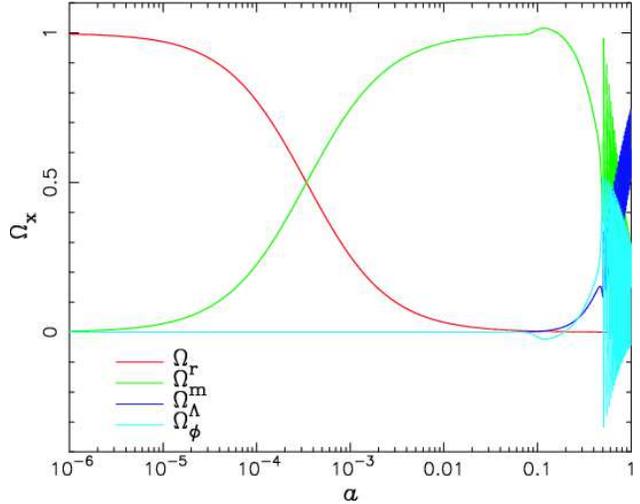}
\caption{Fractional contribution of radiation, matter, $\mit\Lambda$-term, and scalar field $\phi$ to the total density parameter at each epoch.  The total density parameter is always unity because of the flat geometry adopted herein. \label{fig:omega_q12000}}
\end{figure}

 As for the perturbation computations, we make use of the formulation generated by \citet{bac00,per02}. The required perturbation equations can be obtained by substituting 
\begin{equation}
F=\frac{c^4}{8\pi G}-\xi\psi^2,
\end{equation}
\begin{equation}
V=\frac{1}{2}\left(\frac{m_{\phi}c^2}{\hbar}\right)\psi^2\exp\left(-q\frac{4\pi G}{3c^4}\psi^2\right)
\end{equation}
 in their formulae (see Appendix). We have also rewritten the part of the perturbation equations in the CAMB using the quantities appropriate for our model.
 As in \citet{per99}, the adiabatic condition is thereby assumed as the initial condition for perturbations. Since the scalar field is practically non-existent in the early epochs of the proposed models, the effect of $\phi$ can be ignored in comparison with the other components comprising the proposed models, so that the universe can be represented with nothing more than a radiation-dominated ${\mit\Lambda}$CDM model, as is shown in Fig.\ref{fig:omega_q12000}, which indicates the fractional contribution of matter(baryonic matter + CDM), radiation, $\mit\Lambda$-term, and scalar field $\phi$ to the total density parameter, which is unity because we are herein concerned only with the flat geometry. \par
 This diagram shows that radiation is dominant over the other constituents
  for $a \le 3.4\times 10^{-4}$. The values of the model parameters employed here are as follows: $\Omega_{\rm b,0}h_{100}^2=0.024$, $\Omega_{\rm m,0}=0.237$, 
$H_0=72$ km ${\rm s^{-1} Mpc^{-1}}$, $\Omega_{\rm \phi,0}=0.15$, $\Omega_{{\mit \Lambda},0}=0.613$,
 $\xi=-40$, $m_\phi=3.7\times 10^{-31}h_{100}~{\rm eV}$, and $q=12,000$, which characterize the best-fit model proposed herein. 
 Hence, the initial perturbations of the densities of the baryonic matter, CDM, and radiation, as well as those of the metrics, may well be assumed to be given with those derived by \citet{ma95} for the $\mit\Lambda$CDM model. In what follows, we shall discuss the theoretical power spectra of the CMB temperature anisotropy computed for the proposed cosmological model. \par
\subsubsection{Cosmological model with Morikawa's scalar field: $q=0$ case}
\begin{figure}[h!]
\includegraphics[scale=0.44]{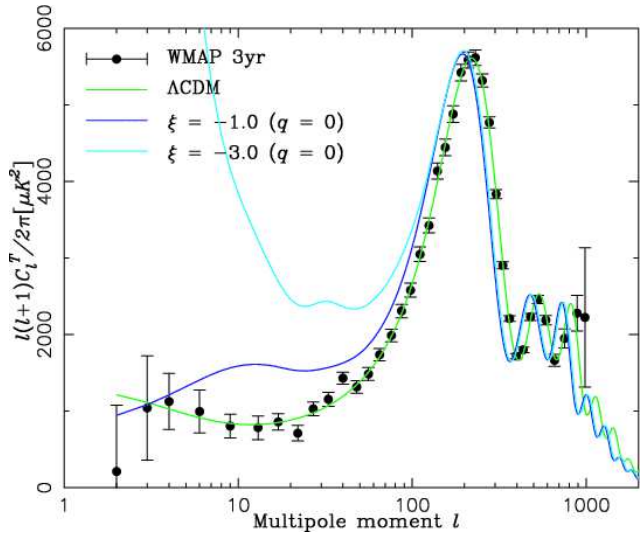}
\caption{Power spectra of the CMB temperature anisotropy computed for 
  the Morikawa scalar field model with  $\xi=-1.0$ and $-3.0$. For the purpose of comparison, the theoretical result obtained by \citet{spe07} for their ${\it \Lambda}$CDM model and the WMAP 3-yr data\citep{hin07} on which it is based are
  also shown. \label{fig:cmb_morikawa}} 
 \end{figure}

Next, let us consider whether we can select an adequate model or models 
 with Morikawa's scalar field from among those that can reproduce the 2dF GRS data
  based on the comparison with the WMAP 3-yr data\citep{hin07}.
  The results of experiments along this line \citep{kaw02,hir06} indicated that it is extremely difficult, if not impossible, to come across such a model because the presence of a scalar field that is large enough to account for the picket-fence structure tends to yield too large a temperature anisotropy of the CMB in the large-scale domain, as shown in Fig.\ref{fig:cmb_morikawa}. 
  Here, we employ $\xi=-1$ and $-3$, the absolute values of which are an order of magnitude smaller than that inferred from the previous analysis of the $N$-$z$ relation of galaxies($\xi=-40$). The values of the other parameters are as follows:
   $\Omega_{\rm m,0}=0.237$, $\Omega_{{\rm b},0}h_{100}^2=0.024$, $\Omega_{\phi,0}=0.15$(hence $\Omega_{{\mit\Lambda},0}=0.613$), $m_\phi=3.7\times 10^{-31}h_{100}~{\rm eV}$, $h_{100}=0.72$ \citep{fre01}, $n_{\rm s}=1$ is the spectral index (Harrison-Zel'dovich spectrum) for the initial condition for the matter density perturbation, $T_{{\rm CMB},0}=2.726$ K \citep{mat99} for the present CMB temperature, and the optical depth of our universe is $\tau=0.089$ for the CMB, \citep{spe07}, which, together with $\xi=-40$,  characterize the best-fit model proposed in the present study. Furthermore, we adopt $Y_{\rm He}=0.24$ for the cosmic abundance(by mass) of He and 3.04 for the number of neutrino species \citep{nag04}.
\par
\par
   Also shown in Fig.\ref{fig:cmb_morikawa} is the theoretical spectrum computed for the $\mit\Lambda$CDM model adopting $\Omega_{\rm b,0}h_{100}^2=0.02229$, $\Omega_{\rm m,0}h_{100}^2=0.1277$, $h_{100}=0.732$, $n_{\rm s}=0.958$, and $\tau=0.089$ derived by the WMAP team \citep{spe07} together with the WMAP 3-yr data.  
Note that the locations of the  peaks of the spectrum computed for the model with the Morikawa scalar field tend to be shifted in the direction of the larger scale side in comparison with those computed for the $\mit\Lambda$CDM model. 
The reason for this is that the horizon scale of the  scalar field model at the time of recombination is greater compared with that of the $\mit\Lambda$CDM model because the effective gravitational constant $G_{\rm eff}$ is smaller in the past. The opposite is true in the case of the scalar field model of \citet{nag04}, for which the value of $G_{\rm eff}$ is larger in the past.\par
Note also that the power spectra obtained for this scalar field model are significantly larger in the large-scale domain than that obtained for the $\mit\Lambda$CDM model.
This may be because the equation of state for the scalar field models fluctuates with time, so that the late-time Sachs-Wolfe effect is enhanced.\par
Hence, we are inevitably led to the conclusion that we cannot possibly hope to reconcile the existence of the picket-fence structure in the $N$-$z$ relation and the spatial power spectrum of the CMB temperature anisotropy within the framework of the cosmological model possessing the Morikawa scalar field. 
 \par
\subsubsection{Cosmological model with a new potential for the scalar field: $q \ne 0$ case}
Note that we have made the scalar field stationary at nearly null level for $a$ roughly less than $5\times10^{-2}$, even in the case of $q=0$ (see Fig.\ref{fig:salgado_m}). It may therefore be inferred that the magnitude of the scalar field must still be too large during the formation of the spatial power spectrum of the CMB. In other words, we might be able to obtain a closer fit to the WMAP data in the large-scale region, if we could retard the growth of the scalar field $\phi$. Based on theoretical considerations and numerical verifications, we developed a simple means by which to achieve this objective, namely, the introduction of a new form of potential $V(\phi)$ for the scalar field, which is proportional to $\phi^2\exp(-q\phi^2)$ with $q$ being constant instead of $\phi^2$ adopted by \citet{mor90,mor91}, as has been indicated in Eqs.(\ref{EqAdot}) and (\ref{EqAdots}).  Moreover, our potential tends to Morikawa's as $q$ approaches $0$.\par
Recall that the different choices for the values of $q$ hardly affect the 
 resulting theoretical curves of the $N$-$z$ relation, as we have 
 demonstrated in Fig. \ref{fig:nz_q}.
 The actual process we used to obtain the final candidate model is 
 a sort of trial-and-error approach in which the quality of fit of the resulting $N$-$z$ curve and the power spectrum of the CMB to the observational data are simultaneously inspected.  
Rather than delineating the technical details here, however, we simply adopt the best-fit model that has been obtained and carry out a sensitivity study for each of the model parameters ($q$, $\xi$, $\Omega_{\phi,0}$, $m_\phi$, $\Omega_{\rm m,0}$, and $\Omega_{\rm b,0}$) in order to illustrate how we can select a particular value for each parameter. Therefore, our procedure does not guarantee that the true best-fit model, if any, can be located. This does not matter, however, because the main objective of the present work is to demonstrate that a candidate model or models that account for both the picket-fence structure and the observed temperature anisotropy of the CMB can be found in a unified manner.\par
We shall employ our best-fit model as the basis for our subsequent analysis, as was done in the $q=0$ case.
\par
\begin{figure}[h!]
\includegraphics[scale=0.44]{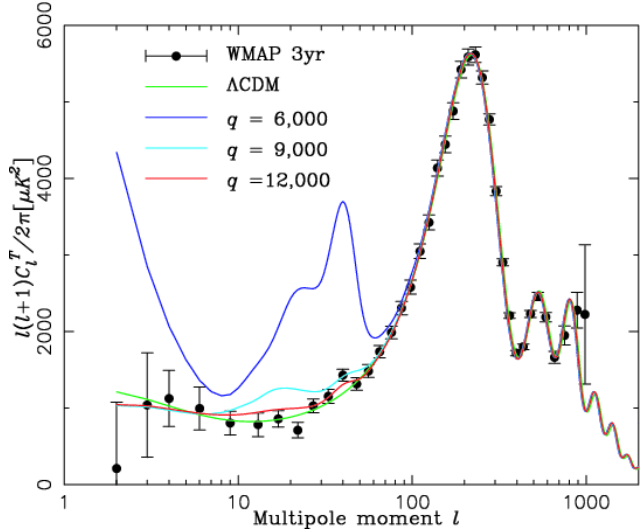}
\caption{Spatial power spectra of the CMB temperature anisotropy computed for our scalar field models with $q=$6,000, 9,000, and 12,000, respectively.
For the purpose of comparison, the spectrum obtained by \citet{spe07} 
for the $\mit\Lambda$CDM model as well as the WMAP 3-yr data are also reproduced. \label{fig:cmb_q_m}}
\end{figure}

 First, let us vary the value of $q$ and compute theoretical power spectra of the CMB temperature anisotropy. The results are shown in Fig. \ref{fig:cmb_q_m} for $q =$ 6,000, 9,000, and 12,000. Also shown are the best-fit curve from the $\mit\Lambda$CDM model \citep{spe07} and the WMAP 3-yr data \citep{hin07}. We are clearly able to produce CMB spatial power spectra, even with the scalar field model, that are at least as good as those obtained by \citet{spe07}. Hence, we adopt $q=12,000$. However, the parameters $q$, $\Omega_{{\rm b},0}$,  $\Omega_{\rm m,0}$, and the optical depth $\tau$ only slightly influence the amplitudes and spatial periods of the picket-fence structure of the $N$-$z$ relation. 
\par
Let us next investigate the sensitivity of the coupling constant $\xi$ to the shape of the spatial power spectrum of the CMB temperature anisotropy.\par
\begin{figure}[h!]
\includegraphics[scale=0.44]{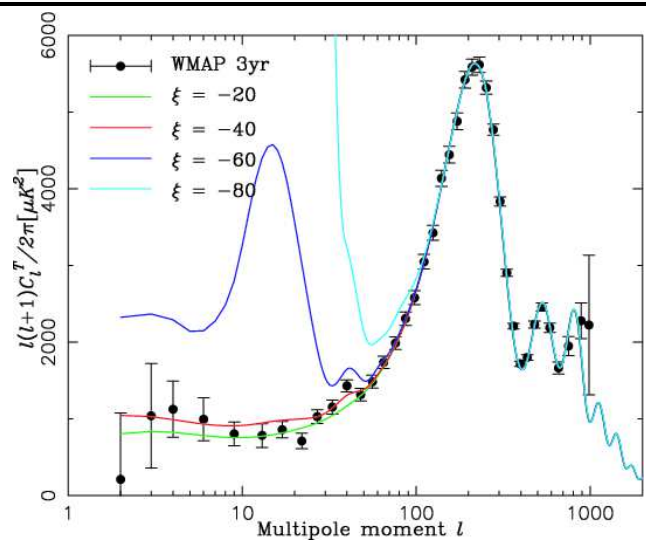}
\caption{Spatial power spectra of the CMB temperature anisotropy obtained for $\xi= -20$, $-40$, $-60$, and $-80$.   The values of the remaining parameters are the same as those employed for Fig. \ref{fig:cmb_q_m}, except that $q=12,000$ is used here.
 Also shown for comparison are the WMAP 3-yr data. \label{fig:cmb_xi}}
\end{figure}

Figure \ref{fig:cmb_xi} shows the theoretical spectra  computed for $\xi=-20$, $-40$, $-60$, and $-80$, but with $q=12,000$ for all of them. The values of the other parameters are the same as those employed for Fig. \ref{fig:cmb_q_m}. The WMAP 3-yr data are also shown in the diagram for comparison. The theoretical spectra with $\xi=-40$ and $-20$ are now in good agreement with the WMAP data. This is in contrast to the case with $q=0$, where $\xi=-3$, or even $-1$, gives rise to an appreciable deviation from the observational data in the large-scale portion, presumably due to the late-time Sachs-Wolfe effect. Taking into consideration the resulting $N$-$z$ relation, we adopt $\xi=-40$ rather than $-20$, as indicated in Fig. \ref{fig:nz_xi}.  \par
\begin{figure}[h!]
\includegraphics[scale=0.44]{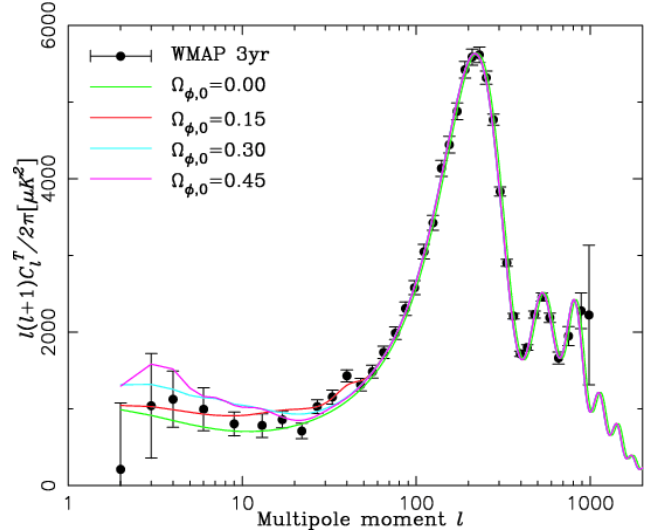}
\caption{Dependence on $\Omega_{\phi,0}$ of the spatial power spectrum of the CMB temperature anisotropy computed with our model. The values of the other parameters are the same as those employed for Fig. \ref{fig:cmb_xi}, except that 
 $\xi=-40$ is used here. For the purpose of comparison, the WMAP 3-yr data are also shown in the diagram. \label{fig:cmb_omegaphi}}
\end{figure}
Figure \ref{fig:cmb_omegaphi} shows the theoretical power spectra of the CMB 
 temperature anisotropy for $\Omega_{\phi,0}=$ $0$, $0.15$, $0.30$, and $0.45$, respectively.  The values of the remaining parameters are the same as those used for Fig. \ref{fig:cmb_xi}, except that we have employed $\xi=-40$ here.
Note that the large-scale portion of the computed spectrum depends rather critically on the value of the density parameter $\Omega_{\phi,0}$ of the scalar
field.  We thus select $\Omega_{\phi,0}=0.15$ as the most appropriate value. \par
\begin{figure}[h!]
\includegraphics[scale=0.44]{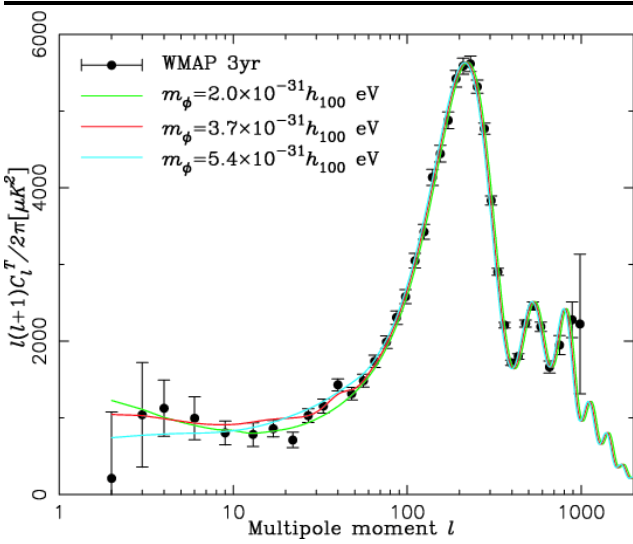}
\caption{Spatial power spectrum of the CMB temperature anisotropy computed for $m_\phi= 2.0\times 10^{-31}h_{100}$~eV, $3.7\times 10^{-31}h_{100}$~eV, and $5.4\times 10^{-31}h_{100}$~eV using our model. The values of the remaining parameters are the same as those for Fig. \ref{fig:cmb_omegaphi}, except that $\Omega_{\phi,0}=0.15$ is employed here. The WMAP data are also shown for the purpose of comparison.
  \label{fig:cmb_ms}}
\end{figure}
\begin{figure}[h!]
\includegraphics[scale=0.44]{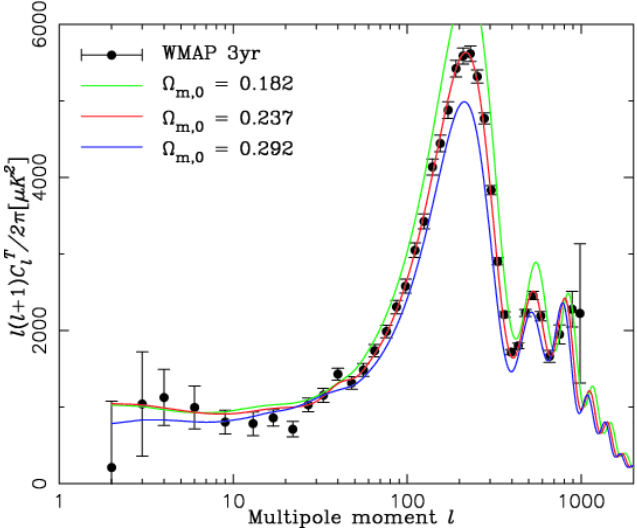}
\caption{Spatial power spectrum of the CMB temperature anisotropy computed 
 for $\Omega_{\rm m,0}= 0.182$, $0.237$, and $0.292$ using our model. 
The values of the remaining parameters are the same as those for Fig. \ref{fig:cmb_ms}, except that $m_{\phi}=3.7\times 10^{-3}h_{100}$ eV is employed here.  For the purpose of comparison, the WMAP 3-yr data are also plotted. \label{fig:omega_m}}
\end{figure} \par
Figure \ref{fig:cmb_ms} shows the CMB power spectra obtained for $m_\phi= 2.0\times 10^{-31}h_{100}$ eV, $3.7\times 10^{-31}h_{100}$ eV, and $5.4\times 10^{-31}h_{100}$ eV, respectively. The values of the other parameters are the same as those for Fig.\ref{fig:cmb_omegaphi}, except that $\Omega_{\phi,0}=0.15$ is employed here. For the purpose of comparison, the WMAP 3-yr data are also plotted in the diagram.  We find it rather difficult to select the best value for $m_\phi$ based solely on the CMB power spectrum. However, the computed $N$-$z$ relation depends strongly on the value of this parameter, as shown in Fig. \ref{fig:nz_ms}, so that we adopt $m_\phi=3.7\times 10^{-31}h_{100}$ eV as the most appropriate value.  \par
Figure \ref{fig:omega_m} shows the CMB power spectra computed for $\Omega_{\rm m,0}=0.182$, $0.237$, and $0.292$. The values of the other parameters are the same as those for Fig. \ref{fig:cmb_ms}, except that $m_\phi=3.7\times 10^{-31}h_{100}$ eV is employed here. The WMAP 3-yr data are also plotted for the purpose of comparison. Note that as the value of $\Omega_{\rm m,0}$ increases, the amplitudes of the peaks, particularly the first and second peaks, of the spectrum tend to decrease significantly. In other words, the value of $\Omega_{\rm m,0}$ has a strong influence on the heights of the peaks, although it hardly affects the spatial periodicity and the amplitudes of the number count variation of the galaxy $N$-$z$ relation. We select $\Omega_{\rm m,0}=0.237$ as the best value based on the comparison of the theoretical curves with the WMAP 3-yr data.\par
\begin{figure}[h!]
\includegraphics[scale=0.44]{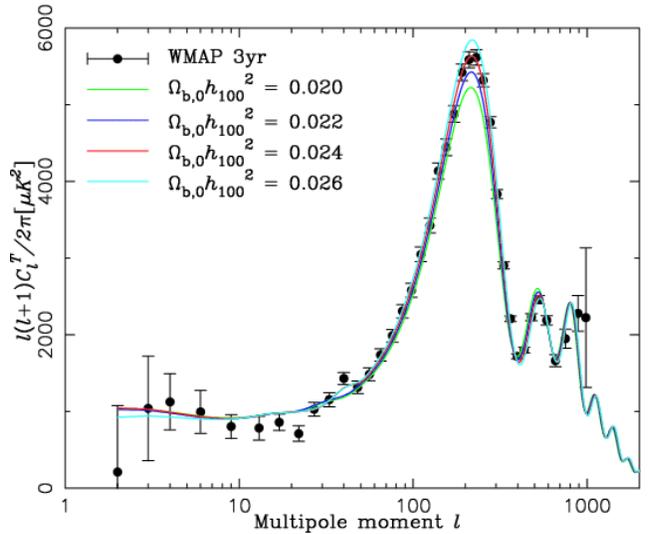}
\caption{Spatial power spectrum  of the CMB temperature anisotropy computed 
 for $\Omega_{\rm b,0}h_{100}^2= 0.020$, $0.022$, $0.024$, and $0.026$ using our model. The values of the remaining parameters are the same as those for Fig. \ref{fig:omega_m}, except that $\Omega_{\rm m,0}=0.237$ is employed here. For the purpose of comparison, the WMAP 3-yr data are also plotted.\label{fig:omega_b}}
\end{figure}

Figure \ref{fig:omega_b} shows the CMB power spectra computed for $\Omega_{\rm b,0}h_{100}^2=0.020$, $0.022$, $0.024$, and $0.026$. The values of the other parameters are the same as those employed for Fig.\ref{fig:omega_m}, except that $\Omega_{\rm m,0}=0.237$ is adopted here. The WMAP 3-yr data are also plotted for comparison. As the value of $\Omega_{\rm b,0}$ increases, the height of the first peak of the theoretical spectrum tends to increase, whereas that of the second peak tends to decrease.  Therefore, we select 0.024 for $\Omega_{\rm b,0}h_{100}^2$ as the best value. \par
\par 
In a manner similar to that described above, we eventually succeeded in producing a model with a non-minimally coupled scalar field that can account not only for the observed spatial periodicity exhibited by the $N$-$z$ relation of the field galaxies but also for the observed spatial power spectrum of the CMB temperature anisotropy. While this model may not be the true best-fit in view of the procedure we have employed to deduce the individual values of the model parameters, we have established the possibility that the observed picket-fence structure of the galaxy $N$-$z$ relation originally discovered by \citet{bro90}, which remains clearly observable in the 2dF data, may be real, rather than an observational fluke. A more elaborate parameter search to find a better model or models using Markov chain Monte Carlo (MCMC) methods (Lewis and Bridle 2002) is currently underway.

\subsection{Type Ia supernovae}
Another important type of observational data that should be considered is  the Hubble diagram of the Ia-type supernovae (SNIa) based on the {\it Supernovae Gold Dataset} compiled by \citet{rie07}, containing a total of 182 objects (119 data points from the previous sample compiled by \citet{rie04}, 16 data points discovered recently using the Hubble Space Telescope (HST), and 47 of 73 distant objects included in the first year release of the Supernova Legacy Survey (SNLS) sample of \citet{ast06}. We would therefore like to test the viability of our best-fit model against this diagram.\par
\begin{figure}[h!]
\includegraphics[scale=0.57]{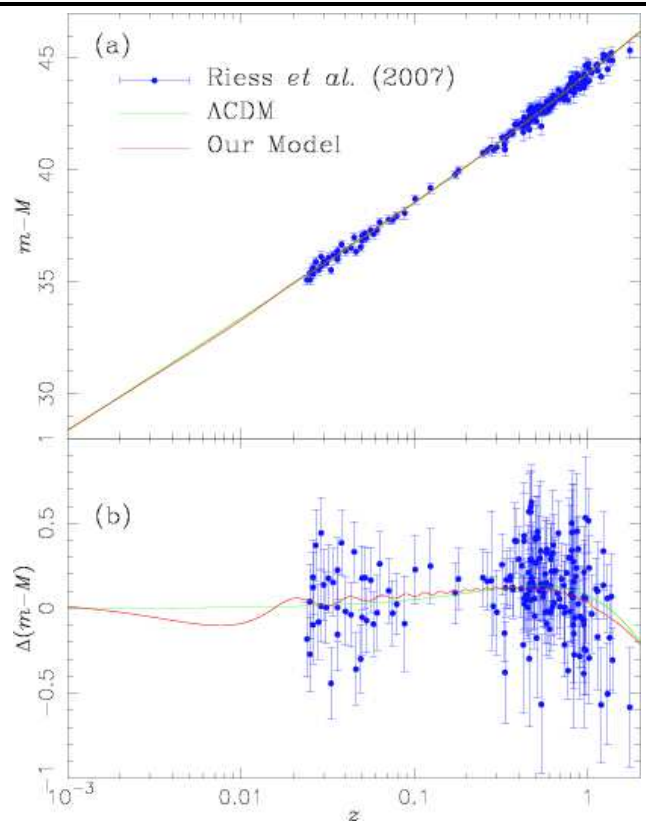}
\caption{(a) Hubble diagram or the distance modulus (ordinate) vs. redshift (abscissa) relation of SNIa's. The discrete points with error bars indicate the observational data taken from the Supernovae Gold Dataset of \citet{rie07}. The solid(red) line curve and the broken(green) line curve denote the theoretical computations obtained for our scalar field model and for the $\mit\Lambda$CDM model of \citet{rie07}, respectively. 
(b) Computed distance moduli and observational data, with the corresponding values given for the constant expansion model being subtracted, plotted as functions of redshift $z$ for better legibility. ($\Omega_{\rm m,0}=0$, $\Omega_{\rm {\mit \Lambda},0}=0$, and $\Omega_{\rm K,0}\equiv -(c/H_0)^2K=1$ corresponding to a negative value of $K$. \label{fig:sn}}
 \end{figure} 
Since we are concerned only with the flat geometry case, we have the following expression for the luminosity distance $D_{\rm L}$ in units of the present-day Hubble radius $c/H_0$:
\begin{equation}
D_{\rm L} = (1+z)\int^z_0 dz^{\prime}\frac{H_0}{H(z)},
\end{equation}
where $H(z)$ is the Hubble parameter. From Eq.(\ref{EqAdot}), $H(z)=(da/dt)/a$ is 
\begin{eqnarray}
H(z)&=&\Bigg[6\xi\phi\phi^{\prime}+\Bigg\{(6\xi\phi\phi^{\prime})^2+(1-6\xi\phi^2)\Bigg.\Bigg. \nonumber \\
    & &\times\Bigg((\phi^{\prime})^2+\left(\frac{m_\phi c^2}{\hbar}\right)^2\phi^2\exp{(-q\phi^2)}\Bigg. \nonumber \\
    & &+H_0^2(\Omega_{m,0}(1+z)^3+\Omega_{r,0}(1+z)^4 \nonumber \\
    & &\Bigg.\left.\Bigg.+\Omega_{{\mit\Lambda},0})\Bigg)\right\}^{1/2}\Bigg]\Bigg/(1-6\xi\phi^2),
\end{eqnarray}
where a primed quantity designates its derivative with respect to 
 the  proper time $t$, viz., $\phi' \equiv {\rm d}\phi/{\rm d}t$.\par
Using this luminosity distance $D_{\rm L}$, we obtain the distance modulus $\mu$:
\begin{equation}
\mu\equiv m-M=5\log_{10}\left(\frac{c}{H_0}D_{\rm L}\right)+25
\end{equation}
 provided that $c/H_0$ is given in units of Mpc. \par
 In Figure \ref{fig:sn}(a), we show the theoretical curve of $\mu$ obtained for our model (solid(red) line) and that obtained for the flat $\mit\Lambda$CDM model with $\Omega_{\rm m,0}=0.29$ of \citet{rie07}(broken(green) line)
as functions of redshift $z$ in the range $[0.001,2]$ together with the 182 data points taken from the Supernova Ia Gold Dataset \citep{rie07}. 
The values of the model parameters employed for our model are the same as those for the best-fit model
  obtained in the previous section:
 viz., $\Omega_{\rm m,0}=0.237$, $\Omega_{\phi,0}=0.15$ (hence, $\Omega_{{\mit \Lambda},0}=0.613$), $\xi=-40$, $m_\phi=3.7\times 10^{-31}h_{100}~{\rm eV}$, and $q=12000$ together with $h_{100}=0.72$(assumed).
  Our theoretical curve is indistinguishable from that of \citet{rie07}. 
In other words, our model is at least as good as the $\mit\Lambda$CDM model of \citet{rie07} so far as the Hubble diagram of SNIa's is concerned. \par
In Figure \ref{fig:sn}(b), the theoretical  distance moduli $\mu$ for our model(solid(red) line curve) and the $\mit\Lambda$CDM model of \citet{rie07}(broken(green) line curve), as well as the observational data, with the corresponding values computed for the constant expansion model universe ($\Omega_{\rm m,0}=0$, $\Omega_{\rm r,0}=0$, $\Omega_{\rm {\mit \Lambda},0}=0$, and $\Omega_{\rm K,0}\equiv -(c/H_0)^2K=1$  corresponding to a negative value of $K$) being subtracted, are shown as functions of redshift $z$ in the range $[0.001, 2]$ for better legibility. Since the oscillation of $\phi$ arises only for $z\le 1$ in our model (see Fig. \ref{fig:model}), our result differs only slightly from that of the $\mit\Lambda$CDM model for $z \simeq 1$ or greater. The $\chi^2$ value for our model and that for the $\mit\Lambda$CDM model are $159.4$ and $160.6$, respectively. However, the difference between the two theoretical curves begins to appear at $z$ less than $\sim 0.02$, and the theoretical curve generated by the proposed model exhibits an oscillatory variation with decreasing $z$. Hence, it is likely that future observations of nearby SNIa's may reveal which type of model is more plausible. 
Note that \citet{laz05} examined 11 types of cosmological models by comparing the resulting theoretical Hubble diagrams with those of a number of SNIa's published by \citet{rie04}, to find  a model employing the equation of state in which the variables oscillate with time to be the most adequate.  \par
\section{Conclusion}
We have succeeded in developing a cosmological model that can explain, in a unified manner and yet within the framework of the flat Robertson-Walker metrics, three types of major cosmological observations, viz., the picket-fence structure or the spatial periodicity of galaxy $N$-$z$ relation, the spatial power spectrum of the CMB temperature anisotropy, and the Hubble diagram of recently observed SNIa's. The picket-fence structure originally discovered by \citet{bro90} in the $N$-$z$ relation of the field galaxies observed through their pencil-beam survey remains distinctly visible  in the 2dF data \citep{col01}, as well as in the SDSS data, although it is less pronounced. Such a feature is difficult to explain with the Robertson-Walker metric cosmology. An ingenious method by which to account for the presence of the picket-fence structure is to introduce a scalar field that is non-minimally coupled with the curvature scalar, as was done by \citet{mor90,mor91}. This enables us to account for the picket-fence structure as an apparent or illusory effect caused by the oscillation of the expansion rate with respect to time. We have therefore investigated if it would be possible to produce a model, accompanied by Morikawa's scalar field, that is capable of accounting simultaneously for the picket-fence structure of the $N$-$z$ relation obtained from the 2dF data and the spatial power spectrum of the CMB temperature anisotropy based on the WMAP 3-yr data. However, the best-fit model deduced from the 2dF data and that derived from the WMAP data have turned out to be mutually exclusive so long as we adhere to the strict Morikawa scalar field.\par
 The most straightforward method by which to avoid this difficulty was to modify the form of the potential for the scalar field. Based on a number of experiments, we have ventured to introduce a new potential function that is proportional to $\phi^2\exp(-q\phi^2)$, with $q$ being constant instead of $\phi^2$ employed by \citet{mor90,mor91}. The effect of $q$ is to retard the time at which $\phi$ starts growing, moving it closer to the present epoch. Furthermore, we need to maintain the scalar field at an extremely small magnitude throughout the epochs from the birth of the CMB to the onset of the growth of $\phi$. 
 In addition, we have found that by solving the evolution equations in the forward direction, starting with such a stationary state at $a=10^{-9}$ and a suitably chosen value for the power index $q$, we can produce a cosmological model in which $\phi$ possesses the desired behavior without requiring any particular fine-tuning  of the initial condition. For $q=12,000$, we are now able to reproduce  the observed power spectrum of the CMB temperature anisotropy as well as the picket-fence structure exhibited by the $N$-$z$ relation by using a model that is characterized by the following values of the model parameters: $\Omega_{\rm m,0}=0.237$, $\Omega_{{\rm b},0}h_{100}^2=0.024$, $\Omega_{\phi,0}=0.15$(hence $\Omega_{{\mit\Lambda},0}=0.613$), $\xi=-40$, $m_\phi=3.7\times 10^{-31}h_{100}~{\rm eV}$, $\phi_0=2.5206\times 10^{-5}$(corresponding to the initial value $\phi_{\rm init}=6.417\times 10^{-31}$), and $n_{\rm s}=1$ (Harrison-Zel'dovich spectrum) together with the assumption that $h_{100}=0.72$ and $T_{{\rm CMB},0}=2.726$ K.\par
\begin{figure}[h!]
\includegraphics[scale=0.38]{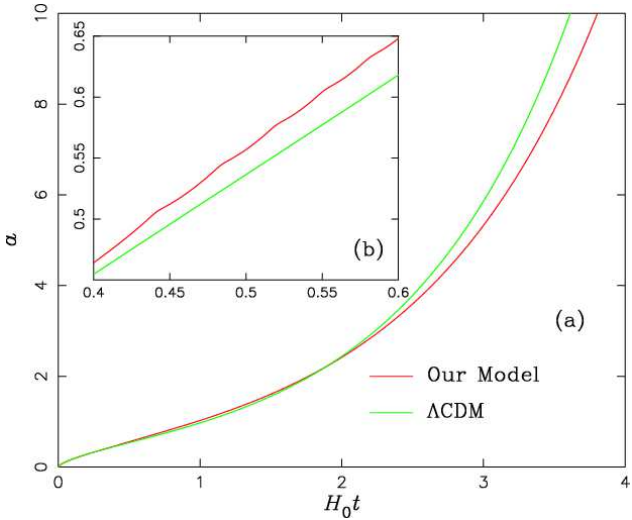}
\caption{(a) Comparison of the time evolution of the scale factor of our model with  that of the $\mit\Lambda$CDM model of \citet{spe07}. The abscissa shows $H_0t$, the proper time $t$ normalized to the Hubble time. (b) Enlargement of the part of Fig. \ref{fig:conclusion_m}(a) in the range of $[0.4, 0.6]$ for $H_0t$. \label{fig:conclusion_m}}
\end{figure}

  Figure \ref{fig:conclusion_m}(a) shows the time evolution of the scale factor of the proposed model in comparison with that of the $\mit\Lambda$CDM model of \citet{spe07}. Note that the rate of expansion of the proposed model is somewhat lower at $H_0t$ above $\simeq 3$, compared with the $\mit\Lambda$CDM model of \citet{spe07} due to the smaller value adopted for  the cosmological constant $\mit\Lambda$ in the proposed model. Figure \ref{fig:conclusion_m}(b) is an enlargement of the area of Fig. \ref{fig:conclusion_m}(a) in the range of $[0.4, 0.6]$ for $H_0t$, which shows a wavy behavior of $a$ for the proposed model because the scalar field is in the phase of a damping oscillation here.  The present age of our model universe turns out to be 13.2 Gyrs, which is somewhat lower than that of the $\mit\Lambda$CDM model of \citet{spe07}. Nevertheless, the lower limit for the cosmic age of 13 Gy set by \citet{kom06} is cleared.
\par
 The proposed model has also been tested against the observed Hubble diagram of the SNIa's created from the Supernovae Ia Gold Dataset of \citet{rie07}. We have found that the proposed model is at least as satisfactory as the $\mit\Lambda$CDM model of \citet{rie07}, although these two models show a rather noticeable difference for $z$ less than approximately 0.02.
\begin{figure}[h!]
\includegraphics[scale=0.37]{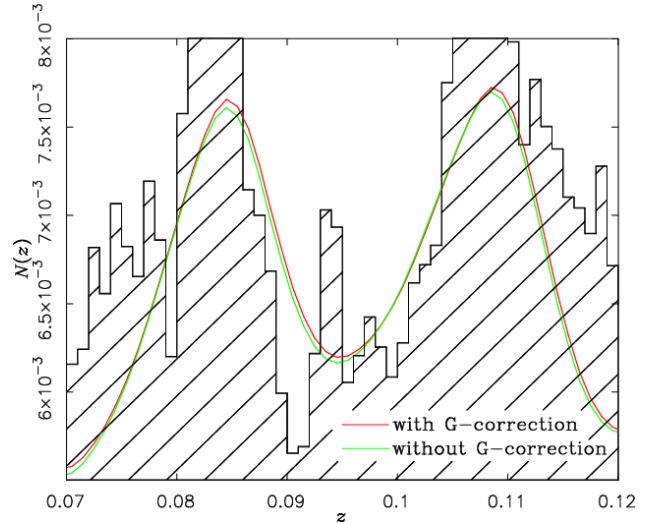}
\caption{Effect of G-correction on the $N$-$z$ relation of galaxies. The abscissa is  the redshift $z$, and the ordinate shows the fractional counts $N$, as in the previous figures. The solid(red) line curve shows the number counts of galaxies taking into consideration the G-correction arising from the temporal variation of the gravitational constant $G$, while the broken(green) line curve indicates the result without  G-correction. The hatched area shows the observational data of the 2dF survey. \label{fig:g_corr}}
\end{figure}
\par
We would now like to turn our attention to the effect of the possible change of the gravitational constant $G$ (see Eq.(\ref{EqGeff})) on the luminosities of galaxies.
\citet{hil90}, for instance, suggested that the spatial periodicity of the galaxy $N$-$z$ relation discovered by \citet{bro90} is due primarily to a periodic  change in the luminosities of galaxies, which might be caused by the variation of the gravitational constant. We have therefore carried out sample computations of the $N$-$z$ relation, taking into account the gravitational correction (G-correction)  $E_{\rm G}(z)$ in obtaining the apparent magnitudes of galaxies employing the proposed best-fit model deduced in the previous sections. According to \citet{han94},  the luminosity $L$ of a galaxy is proportional to $G^{7\sim 8}$, so that we employ the following form for the G-correction \citep{kaw02}:
\begin{equation}
E_{\rm G}(z)=-2.5\log{\left(\frac{G_{\rm eff}(z)}{G_{\rm eff,0}}\right)^{7.5\beta}}, \label{eq:egc}
\end{equation}
where $G_{\rm eff,0}$ is the effective value of the gravitational constant at the present era, while $\beta$ is a positive fudge factor ($\le 1$) to account for the uncertainty involved in the power raised to $G$. For simplicity, however, we adopt $\beta=1$ for our computations. \par
The solid(red) line curve and the broken(green) line curve in  Fig. \ref{fig:g_corr} denote the resulting $N$-$z$ relations for the range $0.07 \le z \le 0.12$ obtained with and  without the G-correction, respectively, whereas the hatched area shows the 2dF data. At $z\simeq 0.1$, for instance, the value of $G_{\rm eff}/G$ is no less than 0.99, as shown in Fig. \ref{fig:model}(d) (see also Eq.(\ref{EqGeff})). Equation (\ref{eq:egc}) predicts that $E_{\rm G}(z)$ shall not exceed 0.08 in magnitude, which in turn indicates that the change in the number counts $N$ caused by the temporal variation of the gravitational constant $G$ should be less than 0.61\%. Hence, we may safely conclude that the primary cause for the picket-fence structure of the spatial periodicity of the galaxy distribution is the oscillatory expansion rate of the universe arising from the oscillation of the non-minimally coupled scalar field, and that the astrophysical effect originating from the variation of $G$ must be at most secondary. 
\par
Finally, the value of the Brans-Dicke parameter $\omega_{\rm BD}$ for our model
 is found to be  
\begin{equation}
\displaystyle{\omega_{\rm BD}={1-6\xi\phi_0^2\over 24\xi^2\phi_0^2}=40,986}
\end{equation}
 which is not inconsistent with the lower limit $\sim 40,000$, referred to as the {\it Cassini bound}, which was set based on recent observations of our solar system \citep{ber03,ber05}.\par






\acknowledgments
The authors would like to thank Y.M. Cho, J.G. Hartnett, K. Ichiki, and M. Morikawa for their helpful comments and discussions. 
We are also grateful to the anonymous referee for the very constructive comments.

\appendix
\section{The Perturbation Equations}
In this Appendix, we present the formulation for the evolution of scalar perturbations of our model. We closely follow the method employed by \citet{bac00,per02} based on the formalism developed by \citet{ma95} to describe the evolution of perturbations in the synchronous gauge. In this type of gauge, the perturbed metric takes the form
\begin{equation}
ds^2=a^2[dt_c^2-(\delta_{ij}+h_{ij})dx^i dx^j],
\end{equation}
where $h_{ij}$ represents the metric perturbations, and its Fourier transform can be written as
\begin{equation}
h_{ij}({\bf x},t_c)=\int d^3ke^{i{\bf k}\cdot{\bf x}}\left[{\bf{\hat{k}_i}}{\bf{\hat{k}}_j}h({\bf k},t_c)+\left({\bf{\hat{k}_i}}{\bf{\hat{k}}_j}-\frac{1}{3}\delta_{ij}\right)6\eta({\bf k},t_c)\right],
\end{equation}
where $h$ denotes the trace of $h_{ij}$, and $\eta$ represents its traceless component. \par
As for the perturbation computations, we employ the formulation reported by \citet{bac00,per02}. By setting 
\begin{equation}
F=\frac{c^4}{8\pi G}-\xi\psi^2,
\end{equation}
\begin{equation}
V=\frac{1}{2}\left(\frac{m_{\phi}c^2}{\hbar}\right)\psi^2\exp\left(-q\frac{4\pi G}{3c^4}\psi^2\right)
\end{equation}
in their formulae, we obtain our perturbation equations. \par
The perturbed density $\delta\rho$, pressure $\delta P$, velocity divergence $\theta$, and shear $\sigma$ obtained based on the formalism of \citet{bac00,per02} are as follows:
\begin{eqnarray}
\delta\rho c^2
                               & = &\frac{1}{1-8\pi G\xi\psi^2/c^4}\left[\delta\rho_{m,r}c^2+\frac{1}{a^2}\dot{\bar{\psi}}\dot{\delta\psi}+6\xi\frac{(\dot{a})^2}{a^4}\bar{\psi}\delta\psi+6\xi\frac{\dot{a}}{a^3}\bar{\psi}\dot{\delta\psi}+6\xi\frac{\dot{a}}{a^3}\delta\psi\dot{\bar\psi}\right. \nonumber \\
& &\left.+\left(\frac{m_\phi c}{\hbar}\right)^2\bar{\psi}\delta\psi(1-q\frac{4\pi G}{3c^4}\bar{\psi}^2)\exp(-q\frac{4\pi G}{3c^4}\bar{\psi}^2)+2\frac{k^2}{a^2}\xi\bar{\psi}\delta\psi+\frac{\dot{h}}{a^2}\xi\bar{\psi}\dot{\bar{\psi}}\right],
\end{eqnarray}

\begin{eqnarray}
\delta P
&=&\frac{1}{1-8\pi G\xi\psi^2/c^4}\left[\delta P_{m,r}+\frac{1}{a^2}\dot{\bar{\psi}}\dot{\delta\psi}-4\xi\frac{\ddot{a}}{a^3}\bar{\psi}\delta\psi+2\xi\frac{(\dot{a})^2}{a^4}\bar{\psi}\delta\psi-2\xi\frac{1}{a^2}\delta\psi\ddot{\bar{\psi}}-2\xi\frac{1}{a^2}\bar{\psi}\ddot{\delta\psi}\right. \nonumber \\
& &-2\xi\frac{\dot{a}}{a^3}\delta\psi\dot{\bar{\psi}}-2\xi\frac{\dot{a}}{a^3}\bar{\psi}\dot{\delta\psi}-4\xi\frac{1}{a^2}\dot{\bar{\psi}}\dot{\delta\psi}-\left(\frac{m_\phi c}{\hbar}\right)^2\bar{\psi}\delta\psi(1-q\frac{4\pi G}{3c^4}\bar{\psi}^2)\exp(-q\frac{4\pi G}{3c^4}\bar{\psi}^2) \nonumber \\
& &\left.-\frac{4}{3}\frac{k^2}{a^2}\xi\bar{\psi}\delta\psi-\frac{2}{3}\frac{\dot{h}}{a^2}\xi\bar{\psi}\dot{\bar{\psi}}\right],
\end{eqnarray}

\begin{equation}
(\rho c^2+P)\theta=\frac{1}{1-8\pi G\xi\psi^2/c^4}\left[(\rho_{m,r}c^2+P_{m,r})\theta_{m,r}+\frac{k^2}{a^2}\left\{\dot{\bar{\psi}}\delta\psi-2\xi(\bar{\psi}\dot{\delta\psi}+\dot{\bar{\psi}}\delta\psi)+2\xi\frac{\dot{a}}{a}\bar{\psi}\delta\psi\right\}\right],
\end{equation}

\begin{equation}
(\rho c^2+P)\sigma = \frac{1}{1-8\pi G\xi\psi^2/c^4}\left[(\rho_{m,r}c^2+P_{m,r})\sigma_{m,r}+\frac{2}{3}\frac{k^2}{a^2}\left\{-2\xi\bar{\psi}\delta\psi+\frac{3}{k^2}(-2\xi\bar{\psi}\dot{\bar{\psi}})(\dot{\eta}+\frac{\dot{h}}{6})\right\}\right],
\end{equation}

where {\it m}  and {\it r} indicate matter and radiation, respectively, $\delta\psi$ is the perturbed part of the scalar field, and $\bar{\psi}$ is the unperturbed part of the scalar field.
\begin{equation}
\psi = \bar{\psi} + \delta\psi.
\end{equation}

The perturbed Klein-Gordon equation is
\begin{equation}
\ddot{\delta\psi}+2\frac{\dot{a}}{a}\dot{\delta\psi}+k^2\delta\psi+6\xi\frac{\ddot{a}}{a}\delta\psi+a^2\xi\bar{\psi}\delta R+a^2(\partial_{\psi}\partial_{\psi}V)\delta\psi+\frac{\dot{h}}{2}\dot{\bar{\psi}}=0.
\end{equation}
Using these equations, the perturbed Einstein equations are
\begin{equation}
\frac{\dot{a}}{a}\dot{h}-2k^2\eta=\frac{8\pi G}{c^4}a^2\delta\rho c^2,
\end{equation}

\begin{equation}
2k^2\dot{\eta}=\frac{8\pi G}{c^4}a^2(\bar{\rho}c^2+\bar{P})\theta,
\end{equation}

\begin{equation}
\frac{1}{3}(2k^2\eta-\ddot{h}-2\frac{\dot{a}}{a}\dot{h})=\frac{8\pi G}{c^4}a^2\delta P,
\end{equation}

\begin{equation}
-\frac{1}{3}(\ddot{h}+6\ddot{\eta}+2\frac{\dot{a}}{a}(\dot{h}+6\dot{\eta})-2k^2\eta)=\frac{8\pi G}{c^4}a^2(\bar{\rho}c^2+\bar{P})\sigma.
\end{equation}

Using these perturbed Einstein equations, we calculate the perturbations on the theoretical CMB anisotropy spectrum.




\clearpage

\end{document}